\documentclass[conference,9pt]{IEEEtran}
\IEEEoverridecommandlockouts

\usepackage{amsmath,amssymb,amsfonts}
\usepackage{algorithmic}
\usepackage{graphicx}
\usepackage{textcomp}
\usepackage{xcolor}
\usepackage{physics}
\usepackage{url}
\usepackage{bbm}
\usepackage[hidelinks]{hyperref}
\usepackage[
    style=ieee,
    citestyle=numeric-comp,
    sorting=none,
]{biblatex}
\usepackage{balance}

\usepackage{mathrsfs}

\def\BibTeX{{\rm B\kern-.05em{\sc i\kern-.025em b}\kern-.08em
    T\kern-.1667em\lower.7ex\hbox{E}\kern-.125emX}}
\newcommand{\hide}[1]{}
\newcommand{\cmt}[1]{}
\newcommand{\etal}{\textit{et al.}}

\begin{document}

\title{Quantum Machine Learning for Finance\\
{\footnotesize \mbox{ }
}}

\author{\IEEEauthorblockN{Marco Pistoia, Syed Farhan Ahmad, Akshay Ajagekar, Alexander Buts, Shouvanik Chakrabarti, Dylan Herman,\\
Shaohan Hu, Andrew Jena, Pierre Minssen, Pradeep Niroula, Arthur Rattew, Yue Sun, Romina Yalovetzky}
\IEEEauthorblockA{\textit{Future Lab for Applied Research and Engineering, JPMorgan Chase Bank, N.A.}}
}

\maketitle

\begin{abstract}
Quantum computers are expected to surpass the computational capabilities of classical computers during this decade, and achieve disruptive impact on numerous industry sectors, particularly finance.  In fact, finance is estimated to be the first industry sector to benefit from Quantum Computing not only in the medium and long terms, but even in the short term.  This review paper presents the state of the art of quantum algorithms for financial applications, with particular focus to those use cases that can be solved via Machine Learning.
\end{abstract}


\section{Introduction}

The computational power of a quantum computer grows exponentially with its number of qubits.  
For this reason, quantum computers are expected to surpass the computational capabilities of classical computers and achieve disruptive impact on numerous industry sectors, 
such as global energy and materials, pharmaceuticals, telecommunication, 
travel and logistics, and finance.  
Finance, in particular, 
is estimated to be the first industry sector to benefit from quantum computing 
not only in the medium and long terms, 
but even in the short term
due to the large number of financial use cases 
that lend themselves to quantum computing 
and their amenability to be solved effectively even in the presence of approximations \cite{mckinsey_quantum}.
This is especially important for taking advantage of today's
Noisy Intermediate-Scale Quantum (or NISQ) devices \cite{preskill2018quantum},
which are characterized by their low quantum bit (or qubit) counts,
short coherence time, and high operation noise.

This review paper presents the state of the art of 
quantum algorithms for financial applications,
focusing in particular on those use cases that can be solved via Machine Learning (ML).
The applicability of ML to finance has become increasingly more significant 
as highly efficient ML algorithms have evolved over time 
to support different data types and scale to larger data sets.
ML operations applicable to finance include regression for asset pricing,
classification for portfolio optimization,
clustering for portfolio risk analysis and stock selection,
generative modeling for market regime identification,
feature extraction for fraud detection,
reinforcement learning for algorithmic trading,
and Natural Language Processing (NLP) for risk assessment,
financial forecasting and accounting and auditing.
Deep learning is often used for image recognition and text classification,
as well as in any use case characterized by large unstructured datasets.

Given the complexity of the algorithms involved,
and the size of the data being analyzed,
ML has been identified 
as one of the most important domains of applicability of Quantum Computing.
This has become even more evident with the discovery 
of new quantum algorithms for linear algebra,
which offer the potential for executing linear-algebra computations on a quantum computer 
more efficiently and accurately than their corresponding classical counterparts \cite{harrow2009linear}.  Under certain conditions, the quantum speedup can be even exponential \cite{Wiebe-exponential-matrix-arithmetics}, modulo some caveats \cite{aaronson2015read}:
\begin{itemize}
\item Efficiently loading classical data onto quantum computers and reading out classical outputs resulting from quantum computations is still the field of ongoing research.  The majority of the quantum algorithms devised so far is based on the existence of a 
Quantum Random Access Memory (QRAM) for accessing the classical data \cite{giovannetti2008quantum}.  The realization of a QRAM has been theoretically proven, but concrete hardware implementations are still undergoing.  Alternatively, classical data can be loaded into a quantum states via specialized circuits \cite{Cortese-loading-classical-data}.
\item It is not always possible 
to apply a quantum linear-algebra algorithm out of the box to solve a specific financial use case; several conditions must be met
and customizations are often necessary to address unique use-case-dependent requirements \cite{aaronson2015read}.
Furthermore, multiple classical and quantum algorithmic components are usually involved 
in the end-to-end solution of a financial use case,
with the potential for any such component 
to become the bottleneck and negate the overall quantum advantage.  The task of computing the quantum speedup of the solution of a specific use case is, therefore, not always intuitive.  
\end{itemize}
As of today, no end-to-end application of quantum ML with exponential speedup over its classical counterpart has been discovered, but several promising directions have been proposed. Meanwhile, a large body of research and engineering work has been successfully dedicated to the realization of quantum algorithms with significant polynomial speedups in their data-processing subroutines, if not in the data loading and output extraction.

\section{Regression}

A central task in supervised learning is \textit{regression}, or the problem of training a simple model to approximate real-valued functions. 
This is an extremely important routine in experimental sciences, and has recently started to be employed on massive datasets. Computationally, this reduces to minimizing a loss function that captures the quality of the fit on training data; the common choices are the $L_\infty$ norm for worst-case error, and the $L_1$ or $L_2$ norms for average-case error. The smoothness of the $L_2$ norm makes it attractive to optimization algorithms, leading to the ubiquity of least squares regressions, where the minimization problem, given $N$ $d$-dimensional feature vectors, can be expressed as follows:
\[
\arg\min_{\vec{\beta}} \sum_{i=1}^{N} w_i \left(y^{(i)} - \vec{\beta}^{T}\vec{x}^{(i)}\right)^{2},
\]
which is a convex quadratic minimization problem and reduces to solving an $N \times d$ system of linear equations. 

A quantum speedup for this task is first considered by Wiebe \etal~\cite{weibe2012data} based on quantum algorithms for solving systems of linear equations. Their algorithm requires access to the entries of the data matrix and weight vector in superposition, and outputs a quantum state that encodes the output in time $\tilde{O}\left(\log(N)s^{3}\kappa^{6}/\epsilon\right)$, 
where $\tilde{O}(\cdot)$ neglects polylogarithmic factors in the complexity.
While the run time of this algorithm is exponentially smaller than classical equivalents, there are two main caveats.
First, the construction of a data structure allowing superposition access to the data points may in general require $O(N \log(N))$ time, via the construction of a QRAM  \cite{giovannetti2008quantum}. Second, $\sim O(d)$ copies of the output state are required to obtain a full description of the output. 

Some attempts have been made to address this issue, albeit with smaller speedups. Wang~\cite{wang2017regression} gives an algorithm for $L_2$ regression that obtains a classical solution in time $\mathrm{poly}(\log(N),d,\kappa,1/\epsilon)$. A second approach is to use algorithms that do not output a full description of $\vec{\beta}$, but rather $\vec{\beta}^T \vec{x}$ for a new data point $\vec{x}$. 
Schuld \etal~\cite{schuld2016regression} obtain such an algorithm given a quantum state encoding of the training data and $\vec{x}$ that takes time $O(\log(N)\kappa^{2}/\epsilon^{3})$. A third approach is to consider special quantum data structures for accessing the training data. These \emph{efficient} QRAM data structures are different from a general QRAM, in that the cost of inserting, updating or deleting a single entry is $O(\mathrm{poly}(\log(n)))$. Such data structures can be useful when there is a large initial corpus of data, but the regression task must be performed repeatedly, and each updates to the data set is small (or constant). The construction of the original dataset thus has cost $O(N \log(N))$, but future upkeep and regression tasks each cost $O(\mathrm{poly}(\log(n,d)))$. Thus, the cost of the initial construction is effectively amortized over the lifetime of the deployment.   This setting can be used to obtain $L_2$-regression algorithms with complexity $\tilde{O}\left(\kappa\mu/\epsilon (\mathrm{poly}(\log(Nd)))\right)$~\cite{chakraborty2019regression,kerenidis2020gradient}. 

There also exist quantum-inspired classical algorithms~\cite{chia2018quantuminspired} based on the last approach that use data structures providing sampling access to the data and obtain algorithms with cost $\tilde{O}\left(\lVert X \rVert_F^{6}\lVert X \rVert^{2}/\epsilon^{4}\right)$.
$L_1$ and $L_\infty$ regression problems are not smooth and are therefore often solved via smooth relaxations that are problem-specific.

A heuristic approach to regression can be used by simply training parameterized quantum circuits (PQC) as function approximators. A common application of classical regression techniques is in learning time series via Recurrent Neural Networks (RNNs) (often using Long Short Term Memory (LSTM) units). These techniques are commonly used to make predictions about evolving processes from historical data. Quantum versions of RNNs have been proposed that use PQCs as the function model \cite{bausch2020recurrent, chen2020quantum, takaki2021temporal} that empirically anticipate improvements in convergence or error rate.\\

With the above discussion on various quantum regression algorithms,
we next look at several financial applications 
that take advantage of these techniques.

\subsection{Asset Pricing}
\textit{Asset Pricing} is the task of assigning prices to various categories of financial instruments, such as stocks, bonds, and derivatives. There are several economic models used to assign these prices based on a sequence of instantaneous (or \textit{spot}) prices, including general equilibrium pricing and arbitrage-free pricing \cite{cochrane2009asset}. The common approach to predicting these spot prices is to model them as functions of simple underlying stochastic processes, such as Brownian or Geometric Brownian motion. Historical financial data can then be used to determine the parameters of these stochastic models. More generally, however, predicting spot prices (as well as many other time-varying quantities of financial interest) can be modeled as a \emph{time series} learning problem. Specifically, given a sequence of historic prices up to time $t$, can accurate predictions be made for prices in the future? Since this reduces to predicting real values based on training data, it is best modeled as a supervised-regression problem. Stochastic pricing models with historically calibrated parameters can be viewed as \textit{ad hoc} solutions based on domain knowledge.

The increasing success of deep RNNs for time-series prediction---especially those that leverage LSTM---has led researchers to consider using these general-purpose algorithms for asset pricing. Gu \etal~\cite{gu2020empirical} and Chen \etal~\cite{chen2021deep} investigate the use of LSTM-based deep-learning methods for asset pricing, obtaining promising results. Specifically, Gu \etal~\cite{gu2020empirical} show that ML forecasts on the S\&P 500 achieve an out-of-sample annualized Sharpe Ratio of 0.77 versus the 0.51 of a buy-and-hold investor. They also find that a value-weighted long-short decile spread strategy based on neural network forecasts of stock prices achieve an annualized Sharpe Ratio of 1.35, nearly double the state-of-the-art classical regression approaches. Chen \etal~further show that refined models based on LSTM forecasts can achieve out-of-sample annualized Sharpe Ratios much larger than naive deep-learning forecasts as well as classical approaches, including the Fama-French five-factor model.

The biggest challenge in deploying deep-learning methods is that training complex neural networks can often be a much more computationally intensive process than the simple parameter calibration required by classical approaches. PQCs may offer advantages over classical variational regression models in terms of expressivity, training complexity and prediction performance. %
In 2020, the use of PQCs to formulate RNNs has been described  \cite{bausch2020recurrent}, along with proposals of quantum LSTM models~\cite{chen2020quantum}. Both approaches show potential empirical improvements over classical neural networks for particular functions, although the applicability to asset pricing is yet to be investigated.

\subsection{Multi-Asset Trend Following Strategies}
Regression models can be used to predict $1$-day returns of a multi-asset class portfolio. Each financial asset class (e.g., equity, bond, cash or commodities) might have different internal dynamics. Nevertheless, a regression model might be able to encompass the global dynamic. An example of global dynamic is the following: if the equity market is bearish, an investor might prefer safer investments, say, bonds, and therefore cause a rise in bond prices. Indeed, there has been a negative correlation between equity and bond since the beginning of the century. However, this correlation was positive between 1970 and 2000 \cite{fan2017equity}. Therefore, the relationship between asset classes evolves and should not be taken as general truth.

To predict returns daily, one can use historical prices from various time points (e.g., 1 month, 3 months, etc.) to introduce trend information in the input data. 
However, this causes an increase in the number of features in the data, and could result in over-fitting. Consequently, one would rather use a Lasso regression \cite{tibshirani1996regression} %
than a vanilla regression. Indeed, by adding an $L_1$ penalty term $\lambda \|\vec{\beta}\|_1$ to the cost function of the $L_2$ regression problem, the model will select a subset of relevant features. As with any regression method, one can treat it as a classification model. The classes here would be buy, hold or sell.
The $L_1$ norm may not be the most attractive regularization term for quantum implementation. Nevertheless, Du \etal~\cite{du2020quantum} provide the closest known implementation with a differentially private Lasso estimator.

\subsection{Implied-Volatility Estimation}
The \textit{Implied volatility} metric captures the financial market's view of the likelihood of changes in a given security's price.
The analysis of volatility is crucial for risk management, portfolio hedging and option pricing. A precise notion of the market's expectation of volatility is required \cite{fengler2002analysis}. Portfolios have a sensitivity with respect to volatility changes. For instance, it has been proven that implied volatilities---such as those of oil, gold and the US stock market---play a role on the returns of the equity sector. Particularly, their impact on the prices and returns of the ten most representative US equity sectors has been quantified \cite{ahmad2021us}. 

A quantum approach for learning implied volatilities has been proposed~\cite{sakuma2020application}.  This uses the deep quantum neural networks firstly introduced by Beer \etal~\cite{beer2020training}.
Given $N$ options, the input data is its strike prices $K_{1}, K_{2}, \ldots, K_{N}$, and the output is the implied volatilities $\sigma_{K_{1}}, \sigma_{K_{2}},\ldots, \sigma_{K_{N}}$. A sigmoid function is used to convert the strike prices to numbers in $[0,1]$, which are then represented as quantum states $\ket{\phi_n^\mathit{in}}$ for $n=1,2,\ldots,N$. The network consists of one input neuron, one output neuron, and one hidden layer with two neurons. The output of the network is a density matrix $\rho_n^\mathit{out} = \ket{\phi_n^\mathit{out}} \bra{\phi_n^\mathit{out}}$. The implied volatility of each of the $N$ options is then calculated using its respective element in the density matrix.

\section{Classification}
 
The goal of \textit{classification} in ML is to predict the labels for new data points using a model that is fit by a labeled dataset. 
Well-known traditional classification algorithms include 
Linear Classification, Nearest Centroid 
and Support Vector Machines (SVMs).
More recently, neural-network-based methods have seen tremendous success.  Once a neural network has had its weights trained via a labeled dataset, it can be used to perform inference on unseen data instances. 
It has been empirically demonstrated 
that neural networks achieve better performance than traditional methods, 
especially on large datasets~\cite{zhang2000neural}.
Neural networks, on the other hand, fall short in terms of transparency and interpretability,
which could be desirable when it comes when making decisions 
or performing tasks that involve handling sensitive information~\cite{johri2020nearest}.  
\paragraph{Linear Classification}
A \textit{linear classifier} allows for classifying an object in a dataset based on the value of a linear combination of that object's characteristics, known as \textit{feature values} and stored in a \textit{feature vector}. 
The common algorithm used for this is the \textit{perceptron} method \cite{minsky2017perceptrons}, which finds a $\gamma$ margin classifier given $n$ $d$-dimensional data points in time $O(nd/\gamma^2)$. 
Quantum algorithms have been shown to be able to provide speedup, bringing the running time down to $O(\sqrt{n}d/\gamma^2)$ \cite{Lloyd2021} or $O(\sqrt{nd}/\gamma)$~\cite{wiebe2016perceptron}.
It was later discovered that the optimal classical algorithm for training classifiers with constant margin runs in $\tilde{O}(n+d)$~\cite{clarkson2012sublinear}, and that a corresponding optimal quantum classification algorithm can bring about quadratic speedup, leading to running time $\tilde{O}(\sqrt{n}+\sqrt{d})$~\cite{li2019sublinear}. 

\paragraph{Distance-based Classifiers}
\textit{Distance-based classifiers} predict the label of a new data point based on its distance to reference points according to some metric.
Typical examples are the  \textit{nearest-centroid} and $k$-\textit{nearest-neighbors} ($k$-NN) classifiers \cite{hastie01statisticallearning}.

The nearest-centroid algorithm is a good baseline classifier that offers interpretable results. This algorithm takes as input a number of labeled data points, where each data point belongs to a specific class. The model fitting consists of computing the \textit{centroids}, which are the barycenters of data points that cluster together in space. Once the centroids are found, a new data point is classified by finding its closest centroid in terms of Euclidean distance.
A quantum version of the nearest centroid algorithm was used to perform classification on the Modified National Institute of Standards and Technology (MNIST) handwritten-digit dataset \cite{johri2020nearest}. The experiments were executed on the IonQ trapped-ion quantum computer.
This approach utilizes novel quantum procedures 
for loading the classical data onto quantum states and estimating distances between these states.
Its accuracy 
matches that of the classical nearest-centroid algorithm. 
The authors, however, 
do not claim any quantum speedup in terms of time complexity.

Another distance-based classifier is the $k$-NN algorithm, which predicts the label of a data point based on a majority vote among $k$ closest training samples according to a metric, such as the Euclidean distance.
Several quantum approaches for $k$-NN have been proposed. In particular, \cite{li2021quantum} utilizes an algorithm for computing Hamming distances in superposition, and the quantum minimum-finding method \cite{durr1996quantum} to find the neighbors with the smallest Hamming distances to a sample.
If feature vectors lie in a low-dimensional space, the algorithm can classify a new sample with worst-case time complexity $O(\sqrt{M}\log(M))$, where $M$ is the training set size. Another quantum $k$-NN approach achieves a query complexity of $O(\sqrt{kM})$ \cite{basheer2020quantum}. The authors design an oracle that encodes the fidelity between two states into a quantum register, which allows for the usage of a quantum algorithm for finding $k$-minima \cite{miyamoto2019quantum}.

\paragraph{Support Vector Machines}
An \textit{SVM}~\cite{boser1992training} consists of solving a convex quadratic optimization problem to find the hyperplane that results in the maximum margin between two classes of data. %
The dual problem  
\begin{equation}
\label{dual_svm}
    \max_{\alpha_i} \sum_{i=1}^N \alpha_i - \frac{1}{2}\sum_{i=1}^N \sum_{k=1}^N \alpha_i \alpha_k y_i y_k K(x_i, x_k) \mid \alpha_i \geq 0\\
\end{equation}
is the one usually solved.
$K : \mathbb{R}^m \times \mathbb{R}^m \mapsto \mathbb{R}$ %
is symmetric and positive semi-definite; a common example is the \emph{Radial Basis Function}. 
$K$ induces the, potentially \emph{non-linear}, feature map $\phi(x) = K(\cdot , x) = K_x$. The Hilbert Space built from such maps is called the \emph{Reproducing Kernel Hilbert Space} (RKHS) with reproducing kernel $K$. 
The optimal classifier in RKHS is one that is a linear combination of $K_x$'s over a subset of the training data~\cite{hastie01statisticallearning}.

One of proposed quantum enhancements to the SVM is based on evidence that universal quantum computation, most likely, cannot be efficiently simulated on a classical computer \cite{quantum_enhanced_feature_spaces_2019}. 
Thus, one should be able to construct a quantum circuit for the map $\vec{x} \mapsto \mathcal{U}_{\Phi{(\vec{x})}}\ket{0}$; $\mathcal{U}_{\Phi{(\vec{x})}}$ is a unitary operation applied to the computational basis state consisting of all qubits in the $\ket{0}$ state, 
such that this operation is not classically feasible. This is called a \emph{quantum feature map} and maps classical data $\vec{x}$ into a quantum Hilbert Space that is exponentially large in the dimension of $\vec{x}$.
A potential quantum kernel is 
\[
K(x_i, x_j)=\lVert\bra{0}\mathcal{U}^\dagger_{\Phi{(\vec{x_i}})}\mathcal{U}_{\Phi{(\vec{x_j})}}\ket{0}\rVert^2
\]
which is symmetric and positive semi-definite. In this case the RKHS is spanned by the functionals, $K(\cdot , x)$, that are constructed from quantum circuits.
The coefficients $\alpha_i$ of the decision function in RKHS can be computed by a convex optimizer running on a classical computer; the quantum computer is used to evaluate the kernel. This hybrid model is called QSVM \cite{quantum_enhanced_feature_spaces_2019}. %
There is potential quantum advantage in the expressability of the feature map, as long as the associated kernel is infeasible for a classical device to compute. This kernel can be computed on a quantum device utilizing either the \emph{destructive SWAP} \cite{Mitarai_2019} or the \emph{controlled-SWAP }\cite{buhrman2001quantum} tests. The latter has better asymptotic complexity, but is not as feasible on NISQ devices. While the former's asymptotic scaling is prohibitive, it can efficiently be implemented on small quantum computers; this allows for experimentation in the near-term.

The class of \emph{Instantaneous Quantum Polynomial} (IQP) circuits 
has been suggested as a potential candidate for $\mathcal{U}_{\Phi{(\vec{x})}}$~\cite{quantum_enhanced_feature_spaces_2019}:
\begin{equation}\label{IQP}
\begin{array}{r@{}l}
\mathcal{U}_{\Phi{(\vec{x})}} &{}= U_{\Phi(\vec{x})} H^{\otimes n}  U_{\Phi(\vec{x})} H^{\otimes n} \\
U_{\Phi(\vec{x})} &{}= \exp\left(i \sum_{S \subseteq [n]} \phi_S(\vec{x}) \prod_{i \in S} Z_i\right)
\end{array},
\end{equation}
where the functions $\phi_S$ represent classical preprocessing.
If the IQP circuit is deep enough, it is believed that computing inner products of states resulting from these embeddings is $\text{\#P}$-Hard \cite{quantum_enhanced_feature_spaces_2019}, thereby potentially out of reach of classical devices. 

A secondary consideration is whether quantum algorithms can be used to accelerate the training of classical SVMs. 
Algorithms of complexity $\tilde{O}(\sqrt{n} + \sqrt{d})$
has been proposed for training kernel classifiers 
and $\ell_2$ margin SVMs~\cite{li2019sublinear}. 
These algorithms are optimal 
and provide a quadratic speedup over corresponding optimal classical algorithms~\cite{clarkson2012sublinear}. 
However these algorithms have complexity polynomial in the inverse of the error $(1/\epsilon)$. 
There exist classical algorithms with $O(\mathrm{poly}(\log(1/\epsilon)))$ complexity using interior-point methods;
Kerenidis \etal~\cite{kerenidis2021quantum} propose a quantum algorithm 
to speed up these methods in terms of the dimension $d$.
While the quantum run time depends on terms that are difficult to bound directly, for random instances,
the quantum algorithm can indeed provide a speedup, leading to a $O(n^{2.59})$ complexity, compared to the $O(n^{3.11})$ classical algorithm's complexity.

In SVM, an optimal hyperplane is obtained that divides the dataset into multiple classes, with a time complexity of $O(\log(1/\epsilon) {\rm poly} (N,M))$, where $N$ represents the feature space dimension, $M$ the number of input points, and $\epsilon$ the accuracy. QSVMs have mathematically been proven to have a run time of 
$O(\log (N M))$~\cite{qsvm_lloyd}.

\paragraph{Variational Quantum Classifiers}

\newcommand{\estim}[1]{\ket{#1}\bra{#1}}

\textit{Variational Quantum Classifiers} (VQCs) are hybrid quantum-classical ML architectures meant for classification tasks that utilize the quantum state space as a feature space to potentially obtain a quantum advantage. A VQC circuit mainly consists of a quantum embedding, a PQC for processing the quantum data, a measurement routine, and a classical optimization loop for updating the parameters of the PQC. First, classical input data $\vec{x}$ is mapped to a quantum state non-linearly using the feature-map circuit, $\mathcal{U}_{\Phi(\vec{x})}$, defined in Equation \ref{IQP}. Applying $\mathcal{U}_{\Phi(\vec{x})}$ to $\ket{0}^n$ results in the state $\ket{\Phi(\vec{x})}$.

Next, a PQC, $W(\vec{\theta})$, is constructed with parameters $\vec{\theta}$. An example of such a PQC is one made from compositions of single qubit rotations and entangling gates.
 PQC architectures have been discussed where descriptors, such as the entangling capability and expressibility, are used to characterize the performance of the PQCs~\cite{expressibility_and_entangling_capability}.

In case of a binary-classification problem, a measurement routine is used to get a binary output. This is accomplished by measuring state $W(\vec{\theta}) \mathcal{U}_{\Phi(\vec{x})} \ket{0}^n$ in the Pauli Z-basis and mapping the output bit-string to a function with binary outcome $f : \{0,1\}^n \rightarrow \{+1,-1\}$. The probability of obtaining an outcome, $y = \pm 1$, is 
\[
p_y(\vec{x}) = \sum_{i \in f^{-1}(y)} \lVert \bra{i} W(\vec{\theta}) \ket{\Phi(\vec{x})} \rVert^2.
\]
We repeat this step for $R$ measurement shots, which gives an empirical distribution, $\hat{p}_y(\vec{x})$. 

Then, a classical cost function is formulated to enable optimizing the parameters $(\vec{\theta},b)$, where $b \in [-1,+1]$ is an added bias parameter. Once the classifier is trained on the training data set using a classical optimizer, the trained circuit can now be used to assign labels to unlabelled data. Several optimizers have been proposed and used, both gradient based, 
such as ADAM and SPSA~\cite{adam_optimizer, li2006simultaneous}, 
and gradient-free ones, such as COBYLA~\cite{optimizers_without_derivatives}. 

VQCs have some limitations, and solving these drawbacks is an active area of research. Barren plateaus occur in optimization algorithms of quantum ML when the parameter search space turns flat once the optimizer is run~\cite{cerezo2020barren,sharma2020trainability}. Architecture design problems, such as choosing the correct cost functions and initializing the parameters, is a very complex process that has not been completely understood yet~\cite{cerezo2021cost}. Additionally, a given variational quantum circuit with fixed form may not be able to capture all of the necessary states in the Hilbert space in its parameterization, and as a result, work on adaptive variational quantum algorithms, such as the Evolutionary Variational Quantum Eigensolver (EVQE), may be applicable to VQC~\cite{rattew2019domain}.

There is a connection between the QSVM and VQC formulations~\cite{quantum_enhanced_feature_spaces_2019, schuld2021supervised, huang2021power}  similar to the connection between classical Neural Networks and SVMs~\cite{jacot2020neural}. There are various discussions on how data encoding affects VQCs \cite{huang2021power, Schuld_2021}, such as repeatedly encoding the inputs \cite{P_rez_Salinas_2020}. In addition, efficient methods were presented for encoding categorical features~\cite{yano2020efficient}. Lastly, there has been research into the expressiveness of PQCs \cite{Schuld_2021,Abbas_2021}.

Support vector machines have been used to predict stock prices for over two decades, but also to predict financial distress and company's credit rating~\cite{svm_in_finance}.\\

Next we look at a few example financial applications where the aforementioned quantum classifications techniques have been applied.

\subsection{Prediction of Binary Options}
SVM can be used to predict the outcome of exotic options. The \textit{double no-touch} is a binary option \cite{nekritin2012binary} with a constant payout and is earned if and only if the underlying asset price remains between a predefined lower and upper bound until expiration. Unlike other options, such as a vanilla call, the payoff is not continuous, but all-or-nothing. Therefore, one can use SVM to separate the two classes corresponding to the binary option outcome. As these classes are not linearly separable, one needs a kernel to predict the outcome. This type of exotic option is often used in foreign exchange. The features selected to train the model could be the average directional index and the ratio between realized volatility over implied volatility.

\subsection{Financial Forecasting}

\textit{Financial forecasting} is a planning tool that helps businesses to adapt to uncertainty based on predictions. Particularly, an algorithm to forecast annual earnings is of interest to any company. Such an algorithm has been proposed \cite{easton2020forecasting} that leverages $k$-NN. It matches a company's recent trend in annual earnings to historical earning sequences of other firms that are similar---known as \textit{neighbor firms}. Some of the features taken into account to find such neighbors include matches based on industry, size and past accruals.

\subsection{Credit Scoring}
\textit{Credit scoring} is a method to evaluate the credit risk of loan applications. It helps credit analysts to decide whether the applicants are worthy of credit. Based on past experience, credit scoring is the prediction of future behavior. An algorithm for this has been proposed using weighted $k$-NN \cite{mukid2018credit}. The credit applicants are classified into one of two groups: a group whose members are likely to repay their debts and another group that should be denied credit because of high likelihood of defaulting.

\section{Clustering}
\textit{Clustering} consists of identifying groups of data points 
that are close to each other according to certain metrics. 
The feature space in which the data is encoded
and the grouping metric are proxies 
for the actual similarities and differences of the data points.
Inspired by quantum mechanics and suitable for high-dimensional data,
\textit{Quantum Clustering} (QC)~\cite{horn2001algorithm} is an algorithm that belongs to the family of density-based clustering algorithms, where clusters are defined by regions of higher density of data points.
The basic idea of QC is to map each data point to a Gaussian distribution centered at that sample. An analytical form computed from the Sch\"{o}dinger equation is used to determine the potential that gives rise to a mixture of these Gaussian as its ground state. The minima in the system's potential energy function are used to identify clusters and are found via gradient-descent methods. 
Other points are assigned to clusters in a similar way.
\textit{Dynamic Quantum Clustering} (DQC)~\cite{weinstein2009dynamic}, an improvement of QC, adopts the time-dependent Schr\"{o}dinger Equation in order to study evolution of quantum states associated with data points and the structure of the potential energy function. 
Being data-agnostic, DQC can be applied in a wide range of fields, 
especially finance, for example on S\&P 500 data~\cite{weinstein2013bigdata}.

Classical-algorithm-inspired quantum-clustering techniques have also been proposed.
For example, $k$-\textit{means} is a well-known classical clustering algorithm
that identifies, among all data points, 
the $k$ most significant clusters and their representative centroids.
Inspired by $k$-means, 
and providing the same robustness guarantees against some level $\delta$ of noise as the classical $\delta$-$k$-means algorithm,
the quantum $q$-\textit{means} algorithm~\cite{kerenidis2018q}
has time complexity that is poly-logarithmic in the
size of the dataset, 
and can be implemented using distance estimation and
quantum matrix multiplication. 
A quantum spectral clustering algorithm for data represented as a
graph has also been proposed~\cite{kerenidis2021quantum}. 
To overcome the potentially huge time/space overhead of loading large datasets onto a quantum device,
\textit{coresets} have been proposed~\cite{tomesh2020coreset}, 
which are small datasets combined with weight functions 
to sufficiently summarize original datasets.
If small enough and still a faithful representation of the original dataset,
a coreset could be used to enable execution on a NISQ computer~\cite{khan2019kmeans, tomesh2020coreset, mendelson2019quantumassisted,aimeur2007quantum}.\\

Next we briefly discuss several use cases of these quantum clustering algorithms in the financial sector.

\subsection{Fraud Detection}
Clustering techniques can be used to perform \textit{anomaly detection} by learning, 
from existing data, the \textit{normal} mode(s), 
and then using this information 
to identify if a new data point is normal or otherwise \textit{anomalous}~\cite{aggarwal2005effective}.
Clustering can improve learning from imbalanced datasets,
which oftentimes is the case for fraud data~\cite{singh2018clustering}.
Clustering can also be combined with additional feature-selection and extraction techniques.
For example, in time series data, a series could be hiking abnormally fast but still stay in a normal value range. 
Adding derivatives into the clustering algorithm can help detecting such an anomaly~\cite{sathyapriya2019cluster}.

\subsection{Stock Selection}
Cluster analysis has also been used by investors for
maximizing profit and minimizing loss.
Stock returns are
likely to be similar in a region thanks to geographic and macroeconomic features. 
Identification of stock clusters allows one to track those with similar
returns but different risks. Once stocks are grouped by cluster analysis, informed investors can use the
output for guidance. They will, for instance, look for same-return stocks and then choose
to minimize risks. Alternatively, they will pick a cluster of same-risk stocks and high return~\cite{da2005stock}.

\subsection{Exchange Rate Regimes}
In 1999, Levy-Yeyati and Sturzenegger~\cite{levy2005classifying} wanted to exhibit the inconsistency between the self-reported \textit{de jure} classification from the International Monetary Fund (IMF) and the actual behavior shown in the data.
In order to overcome bias, the authors proposed to use $k$-means to perform cluster analysis for exchange rate regimes. 
This led to a \textit{de facto} classification, that has then been widely used as well as tested against prior methodologies~\cite{eichengreen2013reliable}.

\subsection{Hedge Fund Clustering}
Due to the variety of hedge fund---and, therefore, investing strategies---it can be hard for investors to classify such investment vehicles. Moreover, hedge funds tend to reveal less information than other type of funds as they do not fall under the same disclosure requirements. To classify hedge funds, predefined classes would not be able to manage correctly future type of hedge funds. Hence, clustering methods, such as $k$-means, have been used to overcome this issue~\cite{das2003hedge}.  The  features considered are based on available characteristics of hedge funds, such as asset classes, size, fees, leverage and liquidity.

\section{Generative Modeling}
A \textit{Generative Model} learns a probability distribution over data~\cite{Goodfellow-et-al-2016}. 
In supervised learning, where the model is provided as a set of input/label pairs $\{(x_i, y_i)\}$, the model learns $P(X, Y)$, the joint probability distribution of inputs and labels~\cite{ng2002discriminative}. In unsupervised learning, these models can be used to generate new data given only samples~\cite{radford2016unsupervised}. Since measuring a quantum state naturally results in a probability distribution over the outcomes, it makes sense to see if quantum computation can be utilized for generative modeling.

\paragraph{Boltzmann Machine}

The \textit{Boltzmann Machine}~\cite{koller2009probabilistic} is defined by a collection of \textit{visible} (observed) and \textit{hidden} (marginalized out) random variables, and an undirected graph of conditional dependencies among them. %
It originates from thermodynamics where the nodes represent a system of correlated classical spins, $s_i$, under an external magnetic field. The classical \textit{Ising Hamiltonian}
\[
\mathcal{H} = -\sum_{i, j} J_{ij}s_is_j -\sum_{i}h_i s_i
\]
represents the energy of the system. Probabilistic inference is performed by sampling from the steady-state distribution---a Gibbs state---over the visible nodes. This is usually done utilizing Markov Chain Monte Carlo (MCMC) methods~\cite{koller2009probabilistic}. In most cases, the graph is restricted to being bipartite to make sampling feasible, resulting in the \textit{Restricted Boltzmann Machine} (RBM)~\cite{Amin_2018}.

To formulate the \emph{quantum} Boltzmann Machine, we quantize the Ising Hamiltonian by making the replacements $s_i \mapsto \sigma^{z}_i$, where $\sigma^{z}_i$ is the Pauli $Z$ spin operator for the $i$-th qubit.
This results in a quantum Hamiltonian, and thus nodes are associated with qubits,
and sampling is performed by projective measurements on the visible qubits.

One potential quantum method to sample from the visible nodes of the Gibbs state is to utilize \emph{Quantum Annealing} (QA)~\cite{farhi2000quantum, Amin_2018, Dixit_2021}. For example, QA can be performed using the D-Wave devices~\cite{Harris_2010}. %

Alternatively, we can prepare the quantum Gibbs state for this system by performing \textit{Imaginary Time Evolution} (ITE)~\cite{Zoufal_2021}. If the initial state is maximally mixed, performing ITE according to a quantum Hamiltonian will result in the associated Gibbs State.
ITE can be performed variationaly, via McLachlan's principle, on a gate-based quantum computer~\cite{Yuan2019theoryofvariational}.
Interestingly, the model introduced by Zoufal \etal~\cite{Zoufal_2021} can be utilized to formulate a Boltzmann Machine without restricted connections that is tractable on a quantum device.

\paragraph{Generative Adversarial Learning}
As another prominent architecture for modeling probability distributions,
\textit{Generative Adversarial Networks} (GANs)~\cite{goodfellow2014generative}, 
operate by simultaneously training a \textit{generator network} $\mathcal{G}_\theta$ 
and a \textit{discriminator network} $\mathcal{D}_\phi$ against each other
through adversarial games,
for which $\mathcal{G}_\theta$ tries to fool $\mathcal{D}_\phi$ by generating fake data samples 
that are non-distinguishable from the ones drawn from the real distribution, 
whereas $\mathcal{D}_\phi$ tries to tell them apart and not be fooled by $\mathcal{G}_\theta$.
Quantum GANs (qGANs)  
have since been proposed~\cite{PhysRevLett.121.040502, PhysRevA.98.012324}
and experimentally tested,
for example, 
on superconducting quantum computers~\cite{Hueaav2761}.
Either of qGAN's generator or discriminator, or both,
can be in the form of quantum circuits.
In addition to the original GAN's cross entropy,
other distance metrics, such as Wasserstein~\cite{chakrabarti2019quantum},
have also been proposed to improve the adversarial training on NISQ devices.

\paragraph{Quantum Born Machine}
Closely related to quantum Boltzmann Machines and qGANs,
\textit{Quantum Born Machines}~\cite{Cheng_2018, coyle2020born}
are another class of methods based on PQCs
that have been studied for performing distributed-learning tasks.
For example, Coyle \etal~\cite{coyle2020born} propose using maximum mean discrepancy, the Stein discrepancy, and the Sinkhorn divergence,
to improve the training of a subclass of quantum circuit Born machines.\\

Having discussed several quantum generative modeling techniques,
we next look at sample use cases in the finance domain
where these techniques can be applied.

\subsection{Fraud Detection}
Quantum versions of the Boltzmann Machine have be utilized for generative-learning and discriminative-learning tasks~\cite{Amin_2018, Dixit_2021}. Specifically for fraud detection, a Variational ITE Boltzmann Machine methodology has been utilized to classify anomalous credit-card transactions~\cite{Zoufal_2021}.
The system Hamiltonian is represented by a sum of Pauli strings whose coefficients are functions of trained parameters and input features. As mentioned earlier, this formulation is not restricted to the Ising Hamiltonian typically utilized by Boltzmann Machines. Predictions are performed by sampling from a single visible qubit indicating whether the transaction was fraudulent.

qGANs were combined with a framework for Generative Adversarial Anomaly detection, AnoGAN~\cite{Herr_2021}. The generator was a PQC; the continuous output from the expectation of Pauli Z operators on each qubit was fed into a classical affine upscaling layer to achieve the full input feature dimension. The goal of the generator was to model the distribution of non-fraudulent transactions.

\subsection{Probability Distribution Preparation}

One crucial step for achieving quantum advantage in many financial applications 
is the efficient preparation of input probability distributions.
qGANs~\cite{zoufal2019quantum, SITU2020193} 
and quantum Born Machines~\cite{Cheng_2018, coyle2020born}
have both been utilized 
to learn PQCs for loading probability distributions.
Upon convergence, the quantum  circuit, 
as an efficient representation of the underlying distribution, 
can for example be used in amplitude estimation
to perform derivative pricing tasks~\cite{stamatopoulos2020option},
with a theoretical quadratic speedup compared to classical Monte Carlo simulations.
Additional techniques have been explored for the general creation of continuous distributions~\cite{haner2018optimizing, grover2002creating}.
Additional techniques exploring the creation of certain families of continuous distributions include the work of Rattew \etal~\cite{rattew2020quantum} for the preparation of normal distributions.

\section{Quantum-Assisted Feature Extraction}
\textit{Feature extraction} refers to the set of techniques used to identify attributes of a dataset potentially helpful in ML tasks such as classification and regression. A quantum algorithm may help in feature extraction by computing properties of the dataset that a classical computer would fail to identify, or would take a very long time to do so. By encoding a data onto a quantum state, we can map a low-dimensional classical data to a much higher dimension in the Hilbert space. The expanded dimensionality of the quantum representation may be used to identify features invisible to a classical algorithm~\cite{schuld2019quantum}. The growing interest in quantum kernels~\cite{chatterjee2016generalized, wang2021towards}, used in conjunction with Support Vector Machines, has also culminated an experimental demonstration~\cite{bartkiewicz2020experimental}.

A widely used algorithm to extract low-dimensional features out of a high-dimensional data is the Principal Component Analysis (PCA). In PCA, a large feature space is analyzed to identify attributes with the highest variance. Classical PCA takes time that is polynomial in the dimension or number of features in the original dataset. If such a classical data is mapped to a quantum density matrix, the quantum version of the algorithm can perform PCA exponentially faster, that is in time polynomial in the logarithm of the dimension~\cite{lloyd2014quantum}. 

Extracting features is particularly challenging while analyzing images where a large number of pixels have to be analyzed to identify image attributes. For these applications, a quantum computer may help in edge detection in images~\cite{zhou2019quantum}.  

In finance, feature extraction may be used in detecting anomalies in transactions. As an example use case, graph theoretic tools are used to study bidding markets to identify colluding communities or cartels~\cite{wachs2019network}. Quantum-aided graph kernel methods~\cite{schuld2019quantum, bai2017quantum} have been proposed to detect non-trivial features, such as \textit{communities}~\cite{shaydulin2019network}, in a graph, which may represent, for instance, a network of financial parties that frequently transact with each other. When working with graph representations of data, we often want to measure the similarity between two graphs. In fact, Gaussian Boson Sampling can be used to check if two graphs are isomorphic to each other~\cite{bradler2021graph}. Moreover, Gaussian Boson Sampling can be used to construct kernel vectors representing the similarity between any two graphs~\cite{Schuld_2021}.

\textit{Feature selection} consists of choosing from a subset of the available features to pass to the model~\cite{hastie01statisticallearning}. This contrasts with methods, such as PCA, that perform a transformation on the features. Feature selection can be formulated as a combinatorial minimization problem with binary decision variables designating whether to select a feature or not. Such binary optimization problems can be solved utilizing QA~\cite{farhi2000quantum}.\\

Below we present examples of how these techniques can be applied to financial use cases.

\subsection{Model Reduction}
PCA is a widely used method for dimensionality reduction
that can be seen from the perspective of singular value decomposition. 
With the matrix decomposition $A= U\Sigma V$, 
where $\Sigma$ is a rectangular diagonal matrix,
the $k$-principal components are the first $k$ columns of $U\Sigma$. 

In 2014, Lloyd \etal~\cite{lloyd2014quantum} described a quantum PCA with exponential speedup over its classical counterpart. This theoretical speedup is realizable under certain conditions as it is based on HHL~\cite{harrow2009linear}. 
The algorithm can be used in finance to ease \textit{model tuning}: as market conditions evolves, models needs to be tuned in order to match the implied volatility---volatility estimated by the model---with the market volatility. By using PCA, one reduces the number of components and, consequently, the number of parameters, thereby easing the model tuning.

For example, in a product based on foreign exchange, the input parameters are various and can range from global market data, such as risk-free interest rate, to asset specifics parameters, such as the spot price. As a consequence, the model tuning becomes computationally expensive due to the high number of inputs.
However, as just the top three principal components can oftentimes explain over 95\% of the output variations, one can tune the model faster and still accurately by using only these three components.

A variation of quantum PCA has been implemented on hardware
~\cite{martin2021toward}
to solve a similar problem by reducing the volatility factor dimension of the Heath-Jarrow-Morton model
~\cite{10.2307/2951677} 
in order to estimate forward rates. 

\subsection{Combinatorial Feature Selection for Credit Score Classification}
As mentioned earlier, feature selection can be cast to a combinatorial optimization problem. %
In the case of supervised learning, it important to select features that are independent and relevant to the learning task. 
More specifically, for classification, the correlation coefficients between label and features can represent the relevance. The correlation matrix of the features can be used to represent the dependence between features. This can be formulated as a Quadratic Unconstrained Binary Optimization (QUBO) problem, where the quadratic terms are the entries of the correlation coefficients between features, and the linear terms are correlations between the features and the label. QA~\cite{farhi2000quantum} can be used to solve the QUBO utilizing heuristics provided by quantum mechanics. This exact formulation, solved with a Quantum Annealer, was applied to reduce the number of features used for assessing the credit worthiness of applicants~\cite{milne2017optimal}.

\section{Reinforcement Learning}

\textit{Reinforcement learning} (RL)~\cite{sutton2018reinforcement} is a ML technique where an agent attempts to learn through interactions with the environment.
Classical RL has demonstrated remarkable capabilities in areas such as video games~\cite{mnih2015human}, board games~\cite{silver2016mastering,silver2017mastering}, robotics~\cite{kober2013reinforcement} and self-driving vehicles~\cite{sallab2017deep}.

Classical RL is often formulated as a Markov Decision Process (MDP). MDPs enable the modeling of environments where actions are non-deterministic---that is, where taking a given action may probabilistically lead to one of multiple possible outcomes. As such, MDPs are useful for modeling many real-world problems where RL agents are exposed to inherent uncertainty. An MDP is characterized by a set of states $s\in S$, a set of actions $a \in A(s)$ available at each state $s$, transition dynamics specifying the probability of obtaining state $s_j$ upon taking action $a$ at state $s_i$, and a reward function $R(s_i, a, s_j)$. Of importance, an agent selects actions according to a \emph{policy} which is maintained as a probability distribution over the actions available at any given state.
The objective of an RL agent is to learn an optimal policy (one which selects actions maximizing the expected cumulative rewards) given that both the transition dynamics and the reward function of the environment are unknown \textit{a priori}.

Utilizing quantum computers to perform RL was first discussed by Dong \etal~in 2005~\cite{dong2005quantum}, with a follow-up in 2008~\cite{dong2008quantum}.
In their approach, the possible actions at any given state in the environment are maintained in a quantum superposition, and amplitude amplification is used to increase the probability of measuring a \textit{good} action at any given state. 
In 2017, Dunjko \etal~published a framework for quantum RL, where they expand upon the amplitude-amplification approach, which assumes access to an oracle representing the environment~\cite{dunjko2017advances}. 
Furthermore, they introduce more general techniques for learning model meta-parameters, and additionally observe that there is significant potential for quantum advantage in luck-favoring task environments (i.e., environments where a lucky agent finds good sequences of actions much sooner than an unlucky agent) following from quantum search-based speedups.
In a 2021 paper, Wang \etal~derive a quantum RL algorithm with quadratic performance improvements in various parameters over corresponding classical algorithms for the evaluation of an optimal policy, state-values, and state-action pair values (q-values) in an MDP~\cite{wang2021quantum}. 
They explain that this work is applicable to any RL problem where the environment may be classically simulated, as a classical circuit implementing the simulator may be efficiently turned into a quantum circuit.
Additionally, recent studies have explored the use of variational PQCs to implement both RL and Deep RL (DRL) in continuous action spaces~\cite{chen2020variational, wu2020quantum}.\\

Next, we present some use cases showing how quantum RL techniques can be utilized in the finance domain.

\subsection{Algorithmic Trading}

The process of executing trades of financial instruments systematically by accounting for market variables with limited or no human intervention is referred to as \textit{algorithmic} or \textit{automated trading}. 
Generally, algorithmic trading is performed by predictions in a supervised manner followed by obtaining optimal trading decisions under uncertainty associated with the corresponding predictions and market volatility. 
RL bypasses the need for predictions by casting algorithmic trading as a sequential decision-making problem wherein trading decisions are obtained directly that maximize the cumulative returns over a finite time horizon~\cite{zhang2020deep}. 
The domain of RL, and more specifically DRL, has demonstrated huge applicability for algorithmic trading~\cite{pricope2021deep}. 
However, such RL approaches for automated trading operate under certain strong assumptions and may benefit from quantum ML techniques for improved time and model complexity.

Algorithmic trading can be cast to a multi-period portfolio-selection problem that involves re-balancing the portions of capital invested in selected assets at each stage. There have been attempts to solve this multi-stage optimization problem with a QA device to obtain an optimal trading trajectory~\cite{rosenberg2016solving}. 
However, this approach does not adopt any RL technique based on policy or value function approximation. Due to the hardware limitations of the current quantum devices, quantum RL approaches have not been directly applied yet to automated trading. Nevertheless, components of algorithmic trading can certainly benefit from quantum advantages offered by quantum RL. 
For instance, the LSTM neural network architecture used as q-value estimator~\cite{li2019deep} could be potentially replaced with quantum LSTM~\cite{chen2020quantum} for improved performance. 
Also, variational quantum circuits~\cite{wu2020quantum} can be used for different DRL components applied to decision-making in algorithmic trading.

\subsection{Market Making}
\textit{Market makers} have an important role in financial markets as they increase the liquidity of exchanges, thereby facilitating transactions and investment~\cite{avellaneda2008high, gueant2013dealing}. 
A market maker is responsible for maintaining a set of sell orders (\textit{asks}) and buy orders (\textit{bids}) at various quantities and prices. When incoming market orders are made on a security held by the market maker, they are required to transact. As such, they inherently assume risk, as a position they are forced to acquire can subsequently depreciate. Market makers profit by taking advantage of the gap, called \textit{spread}, between the lowest ask and highest bid. 
For instance, assuming an incoming market order is made to sell security $X$, the market maker will fulfill the order purchasing it at their bid price. If another market order is immediately made to purchase security $X$, the market maker fulfills the order by selling it at their ask thereby price, obtaining a profit equal to the spread. 

Market making is amenable to quantum RL, where the problem can be modelled with an agent state, taking into account attributes such as inventory and risk-tolerance, and an environment state where the agent only has partial information, which may not necessarily be Markovian~\cite{spooner2018market}.

\section{Natural Language Processing}

\textit{Natural Language Processing} (NLP) is the field concerned with automated text and language analysis. A drawback with most search engines that use classical NLP is that they understand separate words and not a grammatical structure. This has triggered research in distributional compositional semantics (DisCo). A particular DisCo model is the Coecke, Sadrzadeh and Clark (CSC) model~\cite{clark2008compositional, coecke2010mathematical},
based on tensor-product composition inspired by quantum theory. 

In modern classical NLP, the vector space model~\cite{schutze1998automatic} is used to compute the meaning of individual words. 
Given an individual word $w$ in a text, 
its meaning is computed by first setting up basis words 
(i.e., the most common words in the text) 
and then, for each of $w$'s nearby basis words,
counting its frequency through the text.
The proximity of two words is measured by the similarity between them 
and it is calculated, for example, 
using the inner product of their normalized representative vectors. 
These are called %
\textit{distributional methods} 
and cannot be extended to find the meaning of long sentences as two sentences are not typically repeated. 
In contrast, algorithms based on compositional semantics derive the meaning of a sentence from known meanings of component words. 
The DisCo model combines both approaches to introduce grammatical understanding to the composition of word vectors. 

In the CSC model, each grammatical type in the text is assigned a tensor product space based on some grammar (e.g., the Lambek's pregroup grammar~\cite{lambek2008word}). For instance, a transitive verb takes a subject noun as a left argument and an object noun as right argument. 
The meaning of a noun is calculated as in the distributional model; its vector space is denoted as $\mathscr{N}$. 
Therefore, the meaning of a transitive verb is a tensor in the space $\mathscr{N} \otimes \mathscr{L} \otimes \mathscr{N}$, 
where $\mathscr{L}$ is the meaning space for the sentences. 
An important feature of this model is the use of diagrammatic notation for vectors, tensors and linear maps. %
This model has the computational challenge of large tensor product spaces. Even thought there exist classical approaches---such as dimensionality reduction~\cite{polajnar2013learning}---to avoid the calculation of the full tensor product, they make certain assumptions that are not always necessarily met.

The recent development of encoding classical data on quantum hardware using variational PQCs enables quantum NLP to be particularly suitable for NISQ devices. 
In particular, the quantum CSC model can encode linguistic structures faster in comparison to its classical counterpart. 
Its quantum speedup stems from the quantum nearest-neighbor algorithm that is employed for the sentence similarity calculations in the DisCo framework.
If certain conditions are met, for the $N$-dimensional noun meaning space there is a quantum algorithm capable of classifying any CSC model sentence composed of $n$ tensors
into $M$ classes with time $O(\sqrt{MN} \log (M))$,
an improvement over classical methods' $O(NM)$ complexity~\cite{zeng2016quantum}. \\

Below are a few potential applications of the discussed quantum NLP techniques in the financial sector.

\subsection{Risk Assessment}
Banks can quantify the chances of a successful loan payment based on a \textit{credit risk assessment}. 
Usually, the payment capacity is calculated based on previous spending patterns and past loan payment history. 
However, this information is not always available, especially for underbanked applicants.
NLP techniques can be applied to solve this problem, 
by using multiple data points to assess credit risk. 
For instance, NLP can measure attitude and an entrepreneurial mindset in business loans.  
Similarly, it can also point out incoherent data and take it up for more scrutiny. Even more, the subtle aspects, such as the lender's and borrower's emotions during a loan process, can be incorporated with the help of NLP~\cite{purda2015accounting, fisher2016natural}.

\subsection{Financial Forecasting}
\textit{Financial forecasting} is based on many macroeconomic factors, which are unstructured and scattered across different sources. 
This is the reason why NLP techniques are frequently employed~\cite{xing2018natural}. 
For example, NLP has been proposed for classification of news articles as significant or non-significant from the financial point of view~\cite{yildirim2018classification}. 
In addition, \textit{sentiment analysis}, which plays an important role in decision-making by traders, has also been carried out with the help of NLP techniques~\cite{mishev2020evaluation}. 

\subsection{Accounting and Auditing}
Another application of NLP is \textit{accounting and auditing}~\cite{fisher2016natural}, whose objective is the detection and prevention of fraud via evaluation of accounting systems, monitoring of internal controls, assessment of fraud risk, and interpretation of financial data for anomalous trends. NLP has been proposed for the creation of semantic knowledge bases or trees for financial accounting standards. Also auditors can detect anomalies in financial statements by applying NLP techniques.

\section{Conclusion}
In this paper, we presented an introduction of quantum ML techniques and their applications in the financial services sector. We identified seven machine learning tasks, for which several quantum algorithms have been previously proposed in the literature: regression, classification, clustering, generative learning, feature extraction, sequential decision-making, and  Natural Language Processing. 
We analyzed the speedups offered by various quantum ML techniques, and discussed the financial applications that could benefit from such quantum acceleration.
Moreover, where the literature for finance-specific quantum ML techniques remains sparse,  we provide insights into applying state-of-the-art general quantum ML techniques to specific financial use cases. 
Additionally, we consider the realities of implementing quantum computing techniques in the financial sector, for example, by considering the challenges imposed by hardware limitations.
In summary, this article serves as a road map towards enriching the finance industry with quantum ML techniques in the NISQ era and beyond.

\section*{Disclaimer}
This paper was prepared for information purposes by the Future Lab for Applied Research and Engineering (FLARE) group of JPMorgan Chase Bank, N.A..  This paper is not a product of the Research Department of JPMorgan Chase \& Co. or its affiliates.  Neither JPMorgan Chase \& Co. nor any of its affiliates make any explicit or implied representation or warranty and none of them accept any liability in connection with this paper, including, but limited to, the completeness, accuracy, reliability of information contained herein and the potential legal, compliance, tax or accounting effects thereof.  This document is not intended as investment research or investment advice, or a recommendation, offer or solicitation for the purchase or sale of any security, financial instrument, financial product or service, or to be used in any way for evaluating the merits of participating in any transaction.

\newpage

\balance
\printbibliography

\end{document}